\documentclass[manuscript,screen, nonacm]{acmart}

\AtBeginDocument{%
  \providecommand\BibTeX{{%
    \normalfont B\kern-0.5em{\scshape i\kern-0.25em b}\kern-0.8em\TeX}}}

\setcopyright{acmcopyright}
\copyrightyear{2023}
\acmYear{2023}
\acmDOI{XXXXXXX.XXXXXXX}

\acmConference[CSCW '23]{Make sure to enter the correct
  conference title from your rights confirmation emai}{October 13--18,
  2023}{Minneapolis, MN}

\usepackage{amsmath}
\usepackage{mathtools}
\usepackage{amsthm}
\usepackage[capitalize,noabbrev]{cleveref}
\usepackage{enumitem}
\usepackage{multirow}
\usepackage{array}
\usepackage{pifont}
\usepackage{makecell}
\usepackage[autostyle]{csquotes} 
\newcolumntype{x}[1]{>{\centering\let\newline\\\arraybackslash\hspace{0pt}}p{#1}}
\newcolumntype{P}[1]{>{\centering\arraybackslash}p{#1}}

\newlist{rquestion}{enumerate}{2}
\setlist[rquestion,1]{label=RQ\arabic*:,ref=RQ\arabic*}
\setlist[rquestion,2]{label=(\alph*),ref=\thequestionsi(\alph*)}

\newlist{questions}{enumerate}{2}
\setlist[questions,1]{label=RQ:}

\newlist{avenue}{enumerate}{2}
\setlist[avenue,1]{label=RA\arabic*:,ref=RA\arabic*}
\setlist[avenue,2]{label=(\alph*),ref=\thequestionsi(\alph*)}

\usepackage{soul}

\begin{document}

\title[ML-Based Teaching Systems]{ML-Based Teaching Systems: A Conceptual Framework}

\author{Philipp Spitzer}
\email{philipp.spitzer@kit.com}
\orcid{0002-9378-0872}
\affiliation{%
  \institution{Karlsruhe Institute of Technology}
  \streetaddress{Kaiserstra{\ss}e}
  \city{Karlsruhe}
  \state{Baden-W{\"u}rttemberg}
  \country{Germany}
  \postcode{76133}
}
\author{Niklas K{\"u}hl}
\email{kuehl@uni-bayreuth.de}
\affiliation{%
  \institution{University of Bayreuth}
  \streetaddress{Wittelsbacherring}
  \city{Bayreuth}
  \state{Bayern}
  \country{Germany}
  \postcode{95444}
}
\author{Daniel Heinz}
\email{daniel.heinz@kit.edu}
\affiliation{%
  \institution{Karlsruhe Institute of Technology}
  \streetaddress{Kaiserstra{\ss}e}
  \city{Karlsruhe}
  \state{Baden-W{\"u}rttemberg}
  \country{Germany}
  \postcode{76133}
}
\author{Gerhard Satzger}
\email{gerhard.satzger@kit.edu}
\affiliation{%
  \institution{Karlsruhe Institute of Technology}
  \streetaddress{Kaiserstra{\ss}e}
  \city{Karlsruhe}
  \state{Baden-W{\"u}rttemberg}
  \country{Germany}
  \postcode{76133}
}

\renewcommand{\shortauthors}{Spitzer and K{\"u}hl, et al.}

\begin{abstract}

As the shortage of skilled workers continues to be a pressing issue, exacerbated by demographic change, it is becoming a critical challenge for organizations to preserve the knowledge of retiring experts and to pass it on to novices. While this knowledge transfer has traditionally taken place through personal interaction, it lacks scalability and requires significant resources and time. IT-based teaching systems have addressed this scalability issue, but their development is still tedious and time-consuming. In this work, we investigate the potential of machine learning (ML) models to facilitate knowledge transfer in an organizational context, leading to more cost-effective IT-based teaching systems. Through a systematic literature review, we examine key concepts, themes, and dimensions to better understand and design ML-based teaching systems. To do so, we capture and consolidate the capabilities of ML models in IT-based teaching systems, inductively analyze relevant concepts in this context, and determine their interrelationships. We present our findings in the form of a review of the key concepts, themes, and dimensions to understand and inform on ML-based teaching systems. Building on these results, our work contributes to research on computer-supported cooperative work by conceptualizing how ML-based teaching systems can preserve expert knowledge and facilitate its transfer from SMEs to human novices. In this way, we shed light on this emerging subfield of human-computer interaction and serve to build an interdisciplinary research agenda.

\end{abstract}




\maketitle

\section{Introduction}
Imagine this: You play chess games against your best friend, and you get beaten. Every. Single. Time. Internet chess platforms like \textit{Lichess} \citep{lichess} offer a machine learning (ML) based teaching system that can walk you through your lost games, point out mistakes and even teach you what the next optimal move would have been. This way you can gradually establish a better understanding of good and bad decisions and be able to defeat your friend in the next game. Finally, the sweet taste of victory!

The example above highlights how ML-based teaching systems can foster the learning process of inexperienced humans (novices) in specific tasks without the intervention of subject matter experts (SMEs). Besides the path to success in board games, this also shows ways in which ML-based teaching systems can support organizations: To remain successful in the long term, they must continuously train employees \citep{clarizia2021learning}, retain retiring SMEs' knowledge and pass it on to novices so that the know-how about organizational tasks, which is one of the most important assets of an organization, is preserved \citep{hatch2004human, levy2011knowledge}. Demographic effects may amplify this risk of ``knowledge loss'' at times when many SMEs retire at the same time \citep{engbom2019firm}. Additionally, high fluctuations in the employment of younger generations \citep{krahn2015exploring} and the war for talent \citep{kwon2021there} lead to human resource bottlenecks in the expensive transfer of expert knowledge from SMEs to novices \citep{levallet2018organizational} and intensify the need for more autonomous teaching systems \citep{kang2021tech}. 

IT-based teaching systems\footnote{We understand IT-based teaching systems as a configuration of social and technical entities that transfer task-specific knowledge of an SME to a novice through targeted teaching interactions with the system entities.} can pose a solution to these organizational problems. The existing literature has therefore long been concerned with the application of IT in teaching in an organizational context. We broadly summarize these long-standing approaches as ``conventional'' IT-based teaching systems. For instance, computer-supported learning applications are being developed to train employees in problem-solving processes in the manufacturing sector \citep{brown2001using}. The study by \citet{brown2001using} shows that humans who spend more time with computer-supported learning applications have a higher learning gain. E-learning systems have proven effective in teaching humans in the medical sector \citep{shih2013evaluation}. \citet{shih2013evaluation} reveal that humans trained with e-learning systems adopt these systems positively and describe them as their favored learning method for orientation training. Another form of IT-based teaching systems are intelligent tutoring systems which are utilized in various industries \citep{frasson1998designing}, e.g., to conduct employee onboarding processes \citep{akyuz2020effects}. However, building and deploying such systems can be costly and resource-intensive since the knowledge-domain in conventional intelligent tutoring systems must be formalized \citep{gross2015learning}.

\begin{figure}[htbp!]
    \centering{\includegraphics[clip, scale=0.48]{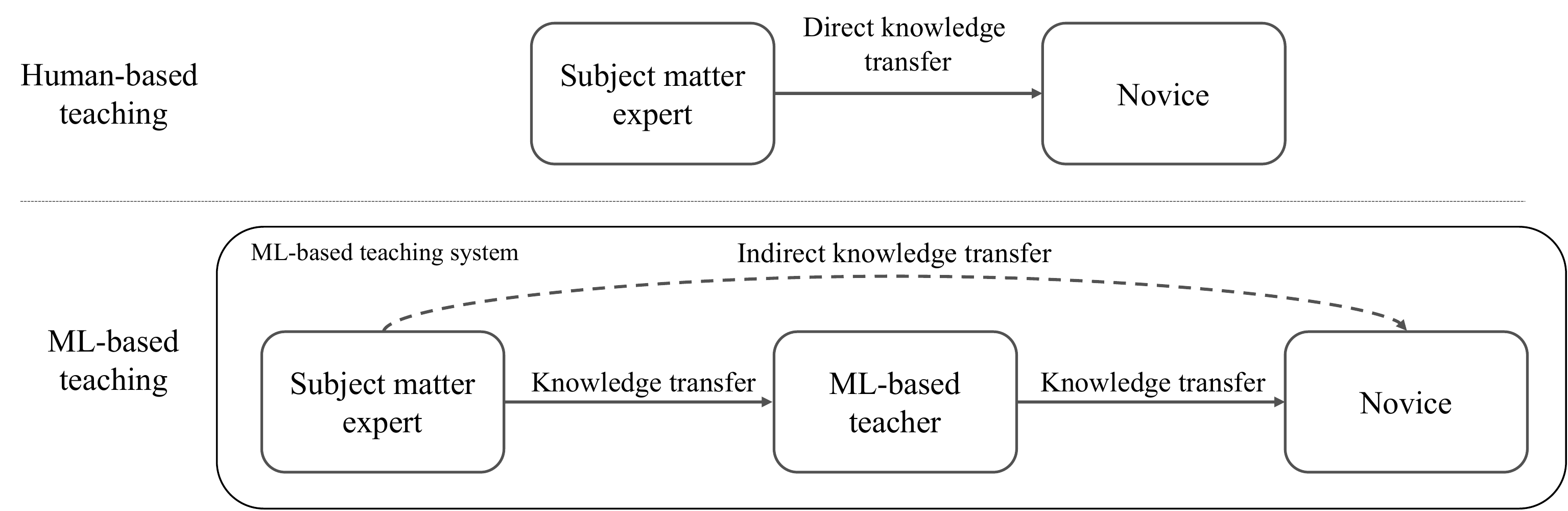}}
	\caption{Human-based teaching versus ML-based teaching.}
        \Description{Human-based teaching versus ML-based teaching.}
	\label{teaching}       
\end{figure}

As a subcategory of IT-based teaching systems, current research in various fields investigates how ML models in instantiations of artificial intelligence (AI) \citep{kuhl2022artificial} can support knowledge transfer from subject matter experts (SMEs) to inexperienced novices. In (\Cref{teaching}) we contrast such ML-based teaching approaches with human-based teaching, where the SME interacts directly with the novice. However, ML-based teaching systems are a currently emerging phenomenon that has not yet been adequately studied, and the existing knowledge in this area is spread across multiple research disciplines without a clear coherence of contributions. In particular, in the field of computer-supported cooperative work and social computing (CSCW), as an interdisciplinary field of human-computer interaction, there is no unified understanding of how ML-based teaching systems as a form of computer systems can support cooperative work. With changing work environments due to demographic effects \citep{engbom2019firm} and digitalization \citep{okkonen2019digitalization} there is a need to synthesize this view.

With this in mind, in this work, we address the need for a common understanding of ML-based teaching systems relevant to both theory and practice. More specifically, we contribute to a better and more integrated understanding of ML-based teaching systems by conducting a review of the recurring patterns of how ML supports knowledge transfer in an organizational context. Hence, we formulate our guiding research question as follows:
\begin{questions}
    \item How can machine learning facilitate knowledge transfer in an organizational context? \label{itm:qwithlabel1}
\end{questions}

To answer this question, we conduct a systematic literature review (SLR) and analyze the existing body of knowledge on ML-based teaching systems \citep{webster2002analyzing}. We use a grounded theory-inspired method for rigorously reviewing literature based on \citet{wolfswinkel2013using} to derive constructs, themes, and interrelationships that define ML-based teaching systems. We present the findings of our review following the procedure of \citet{gioia2013seeking} to illustrate the relations of concepts used in the literature, before synthesizing our findings in the form of a conceptual framework. Finally, we identify emergent interdisciplinary directions for future research. In this context, we scrutinize avenues of relevant CSCW streams to explore how ML-based teaching systems can facilitate knowledge transfer in organizational settings. We thus contribute a consolidated perspective to the emerging field of ML-based teaching systems and respond to the recent call to further examine how ML models can support teaching novices in organizations \citep{seeber2020collaborating}.

The remainder of this paper is structured as follows: First, we outline the foundations and related literature on ML-based teaching systems (\Cref{related_section}, p. \pageref{related_section}) before presenting our methodology for conducting qualitative content analysis (\Cref{methodology_section}, p. \pageref{methodology_section}). We then present the results of our review (\Cref{result_section}, p. \pageref{result_section}) and synthesize them into a conceptual framework (\Cref{framework_section}, p. \pageref{framework_section}). After that, we open up future research avenues and discuss the implications and limitations of our work (\Cref{discussion_section}, p. \pageref{discussion_section}). Finally, we end our article with a brief conclusion (\Cref{conclusion_section}, p. \pageref{conclusion_section}).

\section{Foundations \& Related Work}
\label{related_section}
\subsection{ML-based Teaching Systems}
In recent years there has been an ascent in research on IT-based teaching systems. These systems can provide support in teaching novices through the use of different forms of IT. By providing instructions on specific tasks with accompanying explanations, novices can form new knowledge without direct human interaction, for instance, in role-playing simulations \citep{wang2021practice}. \citet{yang2021can} investigate the minimal amount of explanations needed to teach novices. Hence, such systems can supersede conventional human teachers \citep{anderson1985intelligent}. E-learning systems are an example of IT-based teaching systems that can be utilized for teaching novices, but their effectiveness depends highly on their design, structure, and context \citep{derouin2005learning, fang2022understanding}. However, if set up appropriately, such e-learning systems can be a useful medium for organizations to transfer skills and knowledge to their new employees. As another IT-based teaching approach, related research has focused on adaptive learning systems \citep{huang2012user}. Adaptive learning systems present an approach where ML-based teaching modules are installed to support the operative teaching process, for example, by providing recommendations for an optimal learning path. Such modules can be applied to analyze the learning style of novices and therefore individualize e-learning systems\citep{wakelam2015potential}. 

Unlike such e-learning systems, intelligent tutoring systems typically incorporate a pedagogical model, a student model, a user interface model as well as a domain knowledge model \citep{akkila2019survey}. By identifying the problem-solving state of the novice and using the domain knowledge, such systems can provide instructions based on the novice's needs. For instance, in \citet{wells2021artificial}, the authors describe the deployment of an intelligent tutoring system to train novices on medical tasks. The authors show that such systems can successfully employ ML methods. In \citet{kochmar2022automated}, the authors make use of ML models to select personalized hints for the student. However, \citet{serban2020large} state that the development and installation of intelligent tutoring systems can take up long times. This is because traditional intelligent tutoring systems are based on well-structured domain knowledge. Providing that domain knowledge comes along with a lot of effort \citep{gross2015learning}.

Another teaching concept based on ML is machine teaching (MT). Here, the underlying rationale is to select the optimal teaching set for students, assuming that the teacher knows the decision boundaries \citep{zhu2018overview}. MT is deployed in environments where human teachers train machine learners in an iterative format \citep{wall2019using, ghai2021explainable} or machine teachers teach humans, such as crowd-sourcing workers, to correctly annotate images \citep{wang2020scout}. In such example-based teaching systems, \citet{su2017interpretable} investigate how additional explanations via feature feedback increase the learning progress of novices. However, in MT the machine learner can also collaborate with a human learner in a way that both learners benefit from each other resulting in a more flexible and less expensive setup \citep{nakayama2021crowd}.

A key aspect of research underpins the differentiation of knowledge into its \textit{explicit} and \textit{tacit} forms \citep{polanyi2009tacit, berry1987problem, nonaka1994dynamic}. Various researchers emphasize explicit knowledge as the form humans can articulate with language \citep{nonaka2007knowledge, lam2000tacit}. On the other hand, tacit knowledge is characterized as know-how, with which one can perform particular actions \citep{ryle1945knowing}. It represents human intuition, technical skills, and experience that cannot be expressed and only hardly transferred \citep{nonaka1994dynamic, lam2000tacit, gorman2002types}. Thus, ML-based teaching systems differ to conventional IT-based teaching systems in their ability to capture tacit knowledge of SMEs from data \citep{stein2013machines}. 

Overall, various IT-based teaching systems implement ML models for particular effects. In this work, we consider ML-based teaching systems as teaching systems in which an ML-based model plays a central role to decouple the activities of SMEs and novices.

\subsection{Different Research Perspectives on ML-based Teaching Systems}
In this subsection we provide an overview of different perspectives on ML-based teaching systems to reveal the objectives of various research streams. First, the domain of computer science examines the use of ML-based teaching systems to understand the development and architecture with regards to its technical aspects and effectiveness \citep{wiggins2015javatutor}. \citet{folsom2013tractable}, for instance, make use of an intelligent tutoring system to validate the building of their ML-based teaching system by analyzing the impact of problem characteristics in large-scale problems. 

Second, the human-computer interaction domain investigates ML-based teaching systems to comprehend the properties and principles of interaction designs between human and machine \citep{roessingh2019application, zagalsky2021design}. For instance, \citet{wambsganss2021arguetutor} scrutinize the effect of such teaching systems on the impact on novices. The focus of research is on the interdependency of technological methods such as different ML models and the knowledge building process of novices taking into account cognitive factors of humans \citep{orji2021modelling}. In addition, CSCW research focuses on ML-based teaching systems in work settings. For example, \citet{wang2019human} examine the perception of data scientists towards automated ML tools. Their results show that humans use such systems as teachers for labor-intensive tasks. \citet{sun2019presenters} investigate live-stream teaching with the aid of ML. In their work, they use ML models to study the facial expressions of the audience. The authors find that such systems can assist the teacher in such an online environment. Similarly, in \citet{dillenbourg1997design}, the authors make use of ML models to provide feedback on user interactions to the teacher or to the users themselves.

Third, in the educational research domain, authors shed light on how ML-based teaching systems can be deployed for specific learning needs \citep{troussas2018machine}. They put the focus on utilizing ML models to analyze the learning gains or problematic fields in which novices have issues of comprehending the underlying concepts \citep{piramuthu2005knowledge}. For instance, \citet{yildirim2021artificial} examine the impact of ML models on the efficacy of such teaching systems on novice's ability to form new knowledge.

Fourth, in the realm of psychology, researchers address, among others, how cognitive models of novices affect the teaching process \citep{butler2014integrating}. As pointed out by \citet{evens2006one}, it is also crucial to understand how novices can learn specific concepts and hence, how they establish mental models of underlying principles. Moreover, the interaction of novice and ML model is of interest by taking into account human factors and characteristics \citep{niemi2021ai}.

Fifth, in the field of knowledge management, there is a long tradition when it comes to the transfer of knowledge with support of machine learning \citep{wiig1997supporting}. \citet{alavi2001knowledge} outline ways to store but also transfer knowledge. Whereas one can preserve explicit knowledge through documentation, wikis, or databases, tacit knowledge cannot easily be retained since it is difficult to articulate. \citet{hadjimichael2019toward} give an overview of the different positions that researchers take on the distinction of explicit and tacit knowledge. The authors argue that the interactional perspective of researchers agrees with the idea that tacit knowledge can be converted into data, and used by ML models to learn specific tasks, thus, allowing this tacit knowledge to be transferred. \citet{fenstermacher2005tyranny} suggests utilizing ML models as a medium to formalize tacit knowledge. By doing this, knowledge is preserved within ML models and can be transferred and used to train novices. Finally, \citet{nah2004knowledge} outline the instructions provided by expert systems to enable the distribution of expert knowledge. In \citet{goldstein2018applying}, for instance, the authors make use of ML models to capture the tacit knowledge of an agronomist to recommend proper irrigation plans for orchard plants.

Finally, in the domain of information systems, ML-based teaching systems are examined with a focus on socio-technical aspects of the system. Research in this domain is interested in studying the social aspects of people \citep{mci/Kühl2016} while maintaining a technical perspective on ML \citep{witschel2021dialog}. As an example, \citet{liu2022design} conduct a study to design ML-based teaching systems for specialized contexts during the epidemic period. Furthermore, the information systems domain investigates on design principles and requirements for ML-based teaching systems to successfully teach, for instance, advanced skills to novices \citep{wambsganss2020design}.

In summary, various domains shed light on ML-based teaching systems from different perspectives and a distinct understanding. With this spread interest in the topic it is crucial to assess the capabilities of ML-based teaching systems in detail---to successfully support the knowledge transfer from SMEs to novices. Thus, ML-based teaching systems are a central concept in the field of CSCW, which is an interdisciplinary research area adjacent to the fields outlined in this section. Despite all the work done in CSCW on this topic, there is still a lack of coherent understanding and conceptualization of ML-based teaching systems in organizational settings. Thus, we outline in the following section how we analyze the body of literature on ML-based teaching systems to synthesize common viewpoints and provide a common understanding. One aspect is of increased relevance when it comes to the practical application of the discussed phenomena: Demographic change. The demographic transition of the aging workforce and downsizing strategies of organizations emphasize the need to retain and disseminate the expert knowledge that SMEs possess \citep{levy2011knowledge, schmitt2012don, burmeister2016knowledge}. Knowledge retention within organizations presents a broad research area in which researchers focus on techniques to store and transfer individuals' expertise. For instance, in \citet{levallet2018organizational}, the authors outline several knowledge transfer mechanisms and distinguish between non-IT-based and IT-based ones. The authors highlight that without the installation of proper knowledge transfer mechanisms, organizations can suffer from knowledge loss.

\section{Research Method}
\label{methodology_section}
Our research aims to gain a better overview and conceptual understanding of how ML can facilitate knowledge transfer in an organizational context by reviewing the existing literature. 
To this end, we conduct a broad theorizing review \citep{leidner2018review} as we intend not only to organize and synthesize existing research, but also to bring together different research streams covering the phenomenon of ML-based teaching systems through a (re)conceptualization of this phenomenon.
Our literature review uses a grounded theory-inspired approach as defined by \citet{wolfswinkel2013using} because it facilitates the breaking of established thought patterns. 
Grounded theory, which is typically used in qualitative research \citep{corbin2010basics}, \enquote{ aids in building theory when performing a literature review by focusing on phenomena through a rigorous concept-centric approach.} \citep[p. 52]{wolfswinkel2013using}, as demonstrated by recent reviews on, e.g., digital innovation \citep{hund2021digital}. 
As further established methodological guidance, our research method was informed by \citet{webster2002analyzing}'s suggestions to conduct a concept-centric literature analysis and \citet{gioia2013seeking}'s recommendations to articulate results of inductive data analysis.

\citet{wolfswinkel2013using} introduce five stages as part of their proposed grounded theory literature review method---(1) Define, (2) Search, (3) Select, (4) Analyze, and (5) Present.
Below, we outline how we went through the first four stages of this process before presenting our results (stage 5) in the subsequent section.

\begin{figure}[htbp!]
    \centering{\includegraphics[clip, scale=0.42]{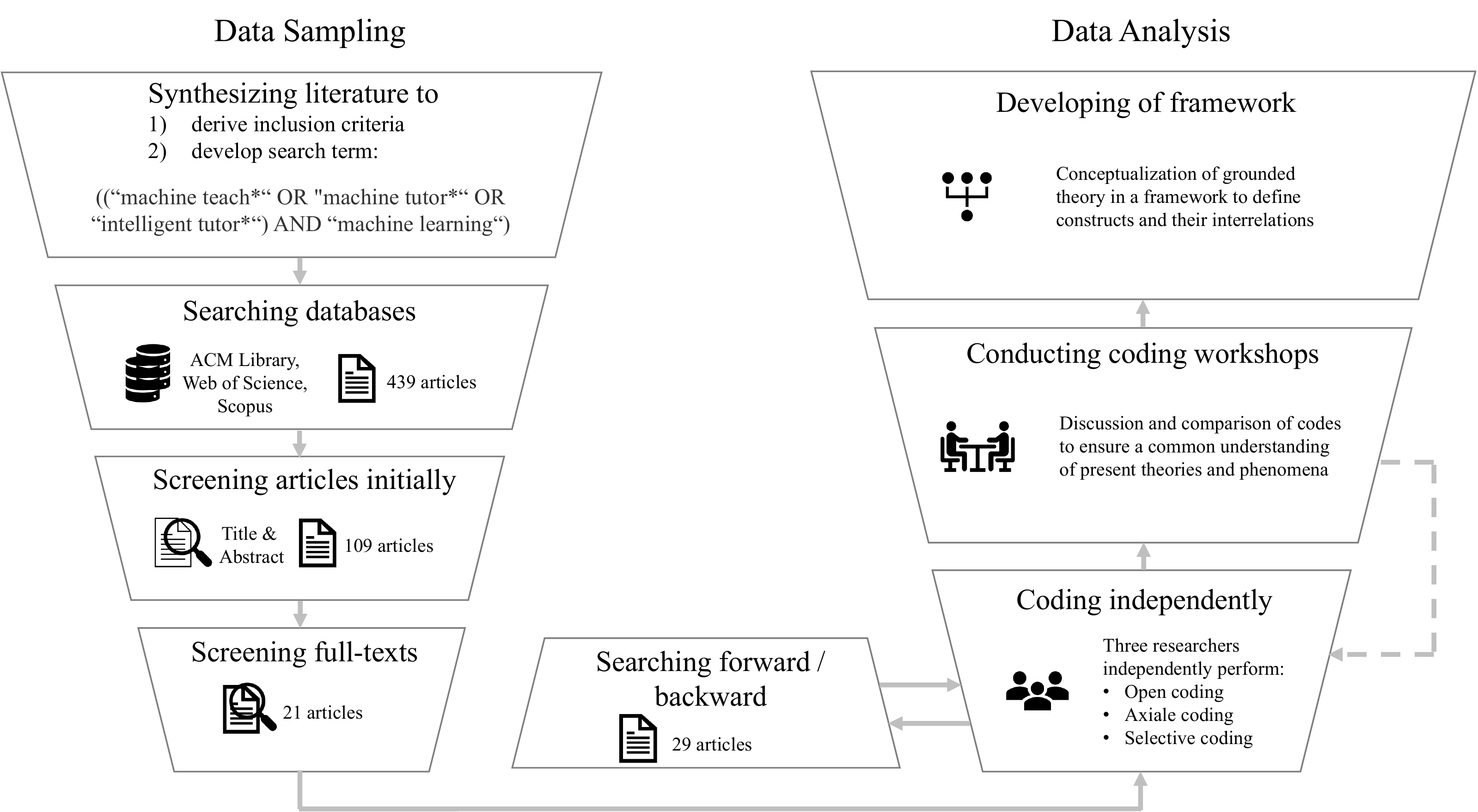}}
	\caption{Inductive analysis through a rigorous literature review and a conceptualization of a framework.}
	\label{slr}       
\end{figure}

\textbf{Define. }In order to grasp the topic of ML-based teaching systems and to develop a basic understanding of the underlying phenomenon, we first individually reviewed an initial sample of articles on ML-based teaching systems and then developed a shared understanding of the phenomenon within the team of authors.
This synthesized understanding allowed us then to define the scope of our review by deriving criteria for inclusion and exclusion and developing a conceptual search term.
In accordance with our research question, we defined criteria to include articles that present a teaching approach in which the knowledge transfer is facilitated by ML models in an organizational context. 
The criteria are based on the following three guiding questions established by the author team:
\begin{itemize}
\item Which capabilities can ML models contribute in teaching systems?
\item Which structural and/or functional roles do ML models take in in teaching systems?
\item What teaching patterns exist in ML-based teaching systems?
\end{itemize}
We explicitly excluded articles in which ML is not primarily used to facilitate the transfer of expert knowledge but rather to support the teaching process in other ways. For instance, teaching systems in which ML methods are used to predict the learning success of students only to optimize the learning path based on conventional approaches are not considered.
Furthermore, we created a search term by decomposing the broad category of ML-based teaching systems into key concepts and deriving synonyms for them in an iterative process of search and refinement. 
The final search term is depicted in \Cref{slr} on the left side.

\textbf{Search. }After completing this process, we conducted a final literature search in the interdisciplinary databases Web of Science, Scopus, and the ACM library in September 2022 to collect potentially relevant articles on ML-based teaching systems. 
Our search in the databases mentioned resulted in a literature sample of 439 articles. 

\begin{figure}[htbp!]
    \centering{\includegraphics[clip, scale=0.6]{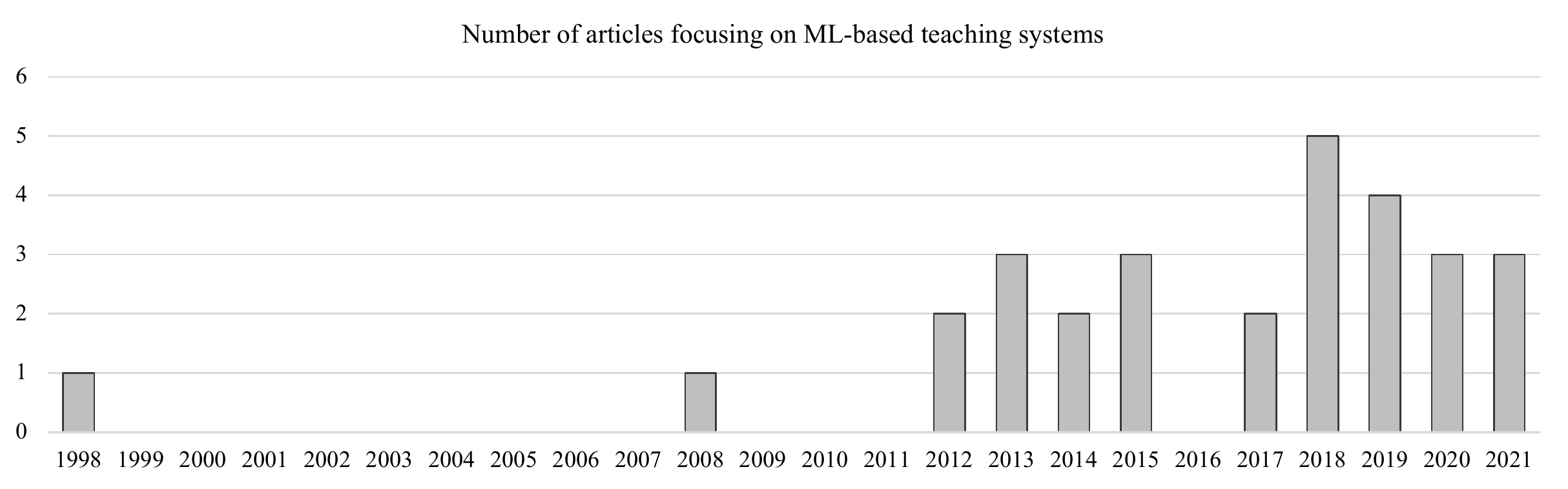}}
	\caption{Publications of articles focusing on ML-based teaching systems.}
	\label{timeline}       
\end{figure}

\textbf{Select. }Based on the previously established inclusion criteria, we initially screened the abstracts and titles of the articles and assessed their relevance to our study. 
This initial screening yielded 109 publications of potential interest. These articles were then screened in full text applying the inclusion and exclusion criteria described above. 
As our reviews followed a grounded theory-inspired approach, we alternated multiple times between data collection and analysis \citep{corbin2010basics}. 
Thus, in analyzing the initial set of articles, we selected a total of 8 additional relevant articles through a forward and backward search and added them to our sample \citep{webster2002analyzing}. 
In this final stage, our literature sample includes 29 articles that examine ML-based teaching systems from a variety of perspectives. \Cref{timeline} illustrates the increasing relevance of the emerging phenomenon of ML-based teaching systems in recent years by showing the number of articles published annually in our sample.

\textbf{Analyze. }In our analysis, we adopted an iterative four-step coding procedure. 
In grounded theory, coding is \enquote{a process of conceptual abstraction by assigning general concepts (codes) to singular incidences in the data} \citep[p. 86]{vollstedt2019introduction}. Throughout this work, the terms ``concept'', ``theme'' and ``dimension'' are used to describe categories in hierarchical order as established in \citep{gioia2013seeking}.
First, we read the articles and immersed ourselves in the literature sample.
During this process, we recorded basic information about each article, namely the outlet, the year of publication, the ML models used, the type of teaching system stated, the theoretical underpinnings, and other information that appeared relevant. 
Second, we retrieved relevant excerpts that aligned with the three guiding questions and used open coding \citep{holton2007coding} to abstract and aggregate core constructs of ML-based teaching systems by annotating the respective excerpts. 
Here, we marked relevant relationships between constructs within each article. 
This procedure was conducted independently by three researchers on a representative sample of the identified articles. 

\begin{figure}[htbp!]
    \centering{\includegraphics[clip, scale=0.47]{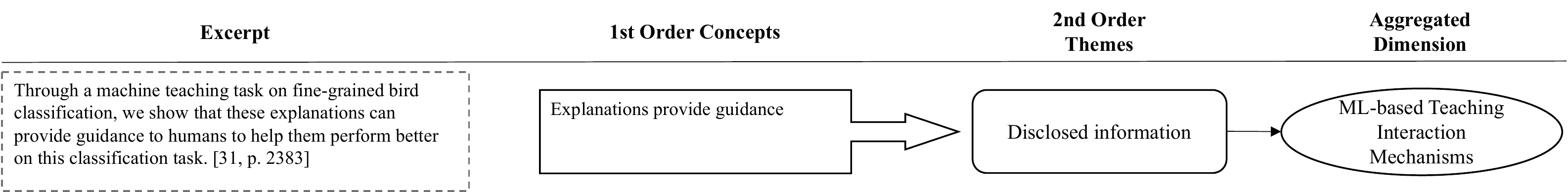}}
	\caption{Example of our Coding Scheme.}
	\label{excerpt}       
\end{figure}

Third, based on the open coding results, we held a workshop to develop a common understanding of the concepts identified in the articles and to ensure a high inter-coder reliability in the ongoing coding process. 
With this shared understanding, we revisited the open codes and conducted a second coding workshop.
In this workshop, we used axial coding to relate the categories to their subcategories and test these relationships against the data, ultimately improving our understanding of the main second-order themes.
These more abstract second-order themes consolidate previously defined concepts. Through this second workshop, we were able to refine our coding structure and develop tentative ideas of the larger narrative within our sample. Inspired by the procedure of \citet{hund2021digital}, the first author then re-coded the remaining articles based on the acquired understanding of the first-order concepts and second-order themes in consultation with the other co-authors. 
In the final step, we conducted a third workshop and used selective coding to further distill the emergent themes into aggregated dimensions as main categories \citep{gioia2013seeking, wolfswinkel2013using}, which were then interrelated to re-conceptualize their connections in form of a conceptual framework.
To illustrate the structure of our results, \Cref{excerpt} shows how we went from an exemplary excerpt to concepts, themes and dimensions (by combining it with other excerpts). In the next two sections, we present the results of this process in form of a review using the derived dimensions, themes, concepts (\Cref{result_section}) and a conceptual framework of ML-based teaching systems (\Cref{framework_section}).

\section{A Review on ML-based Teaching Systems}
\label{result_section}
Overall, analyzing and synthesizing the articles in our literature sample shed light on how ML-based teaching systems are presented in research. The literature contributes to understanding the capabilities of ML models but also the different roles ML models can take on in such teaching systems. In this section, we present the results of our SLR before inductively conceptualizing different concepts of teaching systems in a framework.

As described in \Cref{methodology_section}, we analyzed the content of obtained articles in the SLR according to a concept-centric approach of \citet{webster2002analyzing}. By identifying theoretical concepts, we outline the final 29 articles in a concept matrix (\Cref{appendix}, \Cref{literaturematrix}). We present this concept matrix with a focus on our aggregated dimensions. Since we applied a grounded theory-based approach inspired by \citep{wolfswinkel2013using}, the concept matrix changed during the coding process. We display the final stage of our concept matrix.

\begin{figure}[htbp!]
    \centering{\includegraphics[clip, scale=0.53]{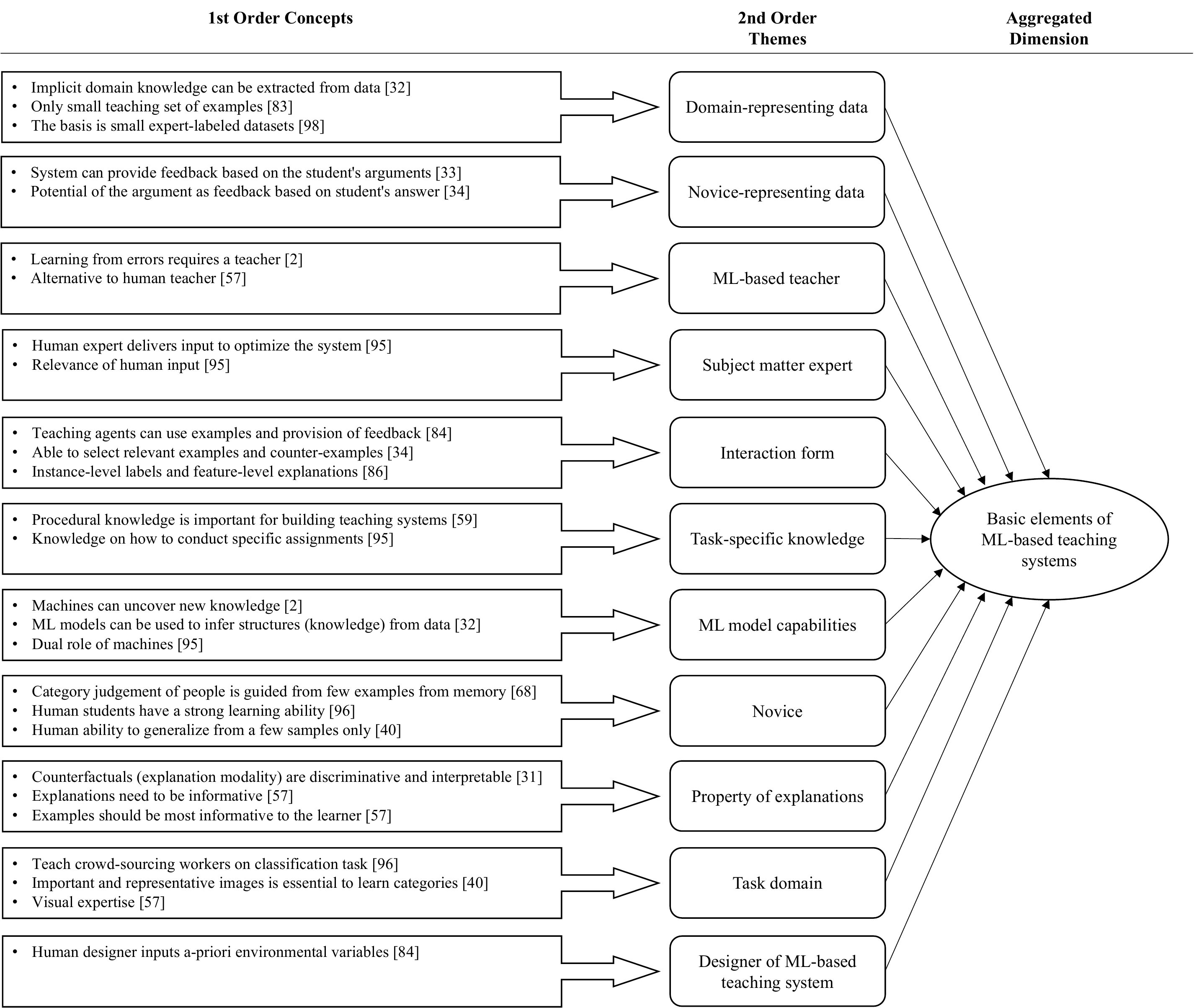}}
	\caption{Data structure of basic elements of ML-based teaching systems.}
	\label{datastructure_be}       
\end{figure}

In our analysis, we derived \textit{first order concepts} from excerpts of the articles collected in the SLR. Based on those concepts, we specified relations between those and defined \textit{second order themes}. Through selective coding we refined the categories and defined \textit{aggregated dimensions} as in \citet{gioia2013seeking}. The resulting data structure can be seen in the figures of the following section (\Cref{datastructure_be}, \Cref{datastructure_ds}, \Cref{datastructure_teim}, \Cref{datastructure_trim}, \Cref{datastructure_rm}). 

In the following subsections, we present the aggregated dimensions \textit{basic elements of ML-based teaching systems}, \textit{design strategies for ML-based teaching systems}, \textit{ML-based teaching interaction mechanisms}, \textit{ML training interaction mechanisms} and \textit{teaching reflection mechanisms} for which we describe the second order themes. Each subsection describes first order concepts relating to the respective theme highlighted in \textbf{bold}. The relationship of second order themes and aggregated dimensions is outlined in \Cref{framework_section}.

\subsection{Basic Elements of ML-based Teaching Systems}
\label{basic_elements_section}
Following our coding methodology in which we iteratively analyzed the underlying concepts, themes, and their interrelations from identified articles \citep{hund2021digital}, we start with outlining the basic elements of ML-based teaching systems (\Cref{datastructure_be}). Basis elements comprise such concepts that are essential elements to set up a ML-based teaching system.

At the core of each teaching system, there is an \textbf{ML-based teacher} embodied by an ML model. They are developed to train novices on specific task domains. Such teachers are the alternative to human teachers \citep{mac2018teaching}. The latter takes over specific functions to facilitate the teaching procedure. The \textbf{interaction form} between ML-based teacher and novice, for instance, can be based on providing examples \citep{guid2019automated}. In this example-based learning set-up, the ML-based teacher can equip the samples with additional information \citep{stein2013machines}, which we define as explanations.

ML models provide different \textbf{capabilities} in teaching systems. One essential ability of such models is to derive structures from data \citep{gross2015learning} and, hence, can be utilized in fields in which the domain knowledge is not well-defined such as learning a new programming language (domain knowledge cannot be formalized per rules). Thus, teaching systems in which the ML model can infer information from an ill-defined knowledge base are easier to deploy than traditional intelligent tutoring systems \citep{gross2015learning}. That is also why ML-based teaching systems can be utilized with less effort \citep{stein2013machines}. In addition, \citet{abdel2020and} characterize ML models as machines that can uncover new knowledge from data. This results from the fact that machines are trained differently than humans and can recognize different patterns and relationships in the domain.

In addition, ML models can take on different roles in teaching systems. In some articles, they initially represent a learner as subject matter experts are training them on a task \citep{stein2013machines}. After reaching a sufficient performance level, they \enquote{turn around} \citep[p. 389]{stein2013machines} and become the teacher themselves. Hence, they depict an intermediary to transfer the \textbf{task-specific knowledge} from SME to novice \citep{wang2019extracting}. Task-specific knowledge depicts the knowledge necessary to conduct a task successfully. In this context, task-specific knowledge defines the knowledge necessary to successfully conduct a task. Another feature that differentiates ML-based teaching systems from conventional teaching by humans is that the former can perform the teaching in an automated fashion \citep{abad2017autonomous}.

\textbf{Novices} can derive concepts based on the examples and explanations provided by the ML-based teacher as they learn \citep{movzina2012goal} and can generalize this knowledge from just a few correct samples compared to ML models \citep{singla2014near, johns2015becoming}. \citet{wang2019extracting} even characterize humans as having profound learning abilities.

In general, ML models utilized in teaching systems that are based on supervised learning methodology are trained on \textbf{domain-representing data}. \textbf{SMEs} can contribute their knowledge by annotating the data. Thus, data has an essential role in the transfer of knowledge from SME to ML-based teachers. Additionally, ML models can extract tacit knowledge from data, which SMEs might not be able to articulate \citep{gross2015learning}. 
By contrast, we distinguish \textbf{novice-representing data} that novices reveal by providing their answer on an assignment. The ML-based teacher can adjust explanations based on the novice's feedback. In \citet{guid2019automated}, the authors outline how novices reveal their answers in the form of arguments.

In addition to providing examples from which novices can defer knowledge, ML-based teachers can provide additional \textbf{explanations} in the form of counterfactual explanations \citep{wang2021machine} or by appending the prediction of the ML model \citep{matsubara2018learning} with the level of uncertainty \citep{abdel2020and}. Various authors state that such explanations shall be informative for the novice to achieve better learning gains \citep{mac2018teaching, cakmak2012algorithmic}. \citet{goyal2019counterfactual} develop an ML-based teaching system for classifying bird species on images. Here, they use explainable artificial intelligence (XAI) methods to create counterfactual examples based on discriminative regions. These are used to make these explanations more interpretable to novices. 

Another first order concept we derived in our qualitative analysis is the \textbf{task domain}. While several authors utilize ML-based teaching systems for classification tasks \citep{johns2015becoming, mac2018teaching, wang2021gradient} to teach visual categories, for instance, crowd-sourcing workers, \citet{stein2013machines} deploy such a system to teach humans psychomotor skills. In \citet{cakmak2012algorithmic}, the authors present a reinforcement learning-based teaching system that explains the optimal teaching strategy to humans. The design of an ML-based teaching system can be controlled by a \textbf{human designer} who can select a priori input variables \citep{stein2013machines}.

\subsection{Design Strategies for ML-based Teaching Systems}
\begin{figure}[htbp!]
    \centering{\includegraphics[clip, scale=0.53]{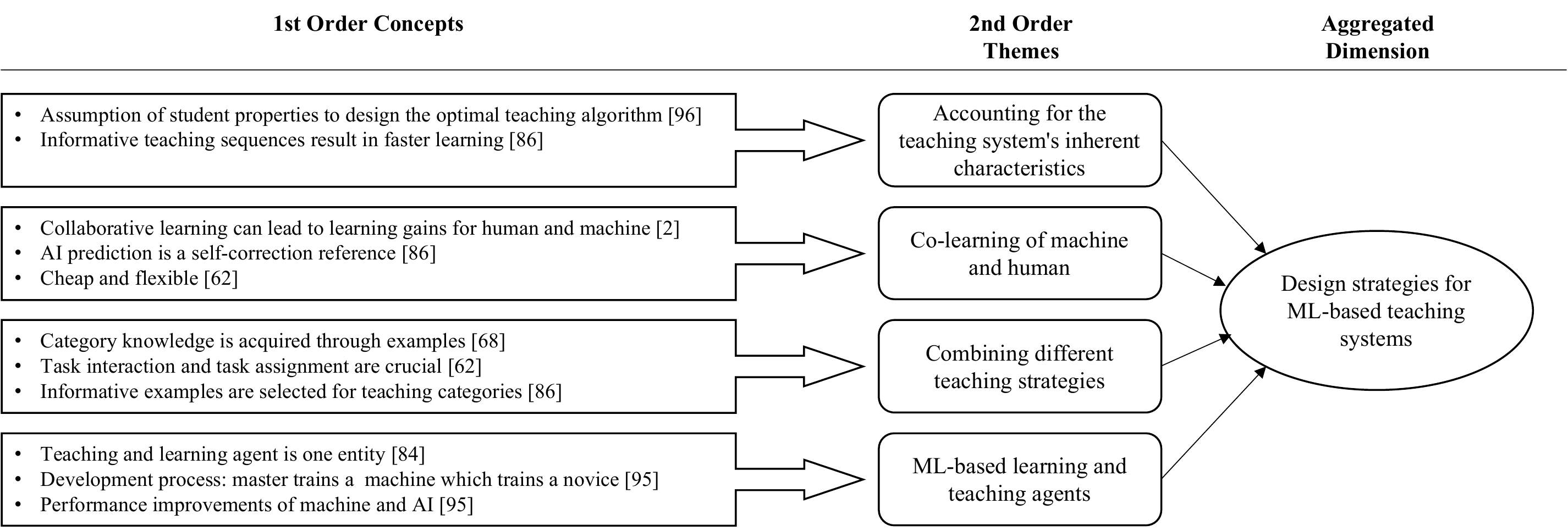}}
	\caption{Data structure of design strategies for ML-based teaching systems.}
	\label{datastructure_ds}       
\end{figure}

The articles in our literature sample present teaching systems that vary in their teaching design, which we refer to as design strategies for ML-based teaching systems. Hence, we distinguish these teaching systems by their \textbf{inherent characteristics} and thereby account for basic patterns in these systems (\Cref{datastructure_ds}). The teaching system, for instance, can be based on an assumption of novice properties to design the optimal teaching procedure \citep{wang2021gradient}. \citet{zhang2020interactive} emphasize reaching a cognitive ability improvement of novices (in their case, crowd-sourcing workers) by utilizing a psychological model-based method. A similar approach is used by \citet{basu2013teaching}. In their work, the authors use mechanisms derived from cognitive science. Based on these models, the teaching system selects examples presented to the novices.

A different design strategy constitutes the mutual learning procedure of novice and ML model, called \textbf{co-learning}. In this collaborative learning setup, humans and machines simultaneously achieve learning gains. Based on task results, data samples are distributed between the human learner and the ML model. Once a training sample is labeled with high confidence, it can be assigned to the other agent \citep{nakayama2021crowd}. This leads to constant learning gains for both agents.

While co-learning as ML-based teaching design can lead to a cheap and flexible workflow \citep{nakayama2021crowd}, designing \textbf{ML-based learning and teaching agents} (LATA) can lead to time and effort reduction \citep{wang2019extracting}. This is because these teaching systems are designed to have the ML model first be trained by an SME and then \enquote{turn around} \citep[p.389]{stein2013machines} and teach a novice. Since such a teaching system is based on observational learning (training the ML model on data samples), one advantage over traditional intelligent tutoring systems is that LATA systems are suitable for domains in which the domain knowledge is ill-structured, e.g., learning how to steer a crane \citep{stein2013machines}. Moreover, LATA agents can also be employed to teach novices skills in specific fields, as \citet{wang2019extracting} illustrate in their article on the example of training humans to play drums.

Overall, we identified that teaching systems can underlie \textbf{combinations of different teaching designs}. Multiple articles describe MT systems in which the objective is to generate an optimal teaching sample presented to novices \citep{johns2015becoming, mac2018teaching, wang2021gradient}. Such sample selection strategies can be supported by giving additional feedback \citep{guid2019automated}. In \citep{abdel2020and}, the authors describe a human-computer collaboration design in which novices are taught. This design combines co-learning and LATA.

\subsection{ML-based Teaching Interaction Mechanisms}
\begin{figure}[htbp!]
    \centering{\includegraphics[clip, scale=0.53]{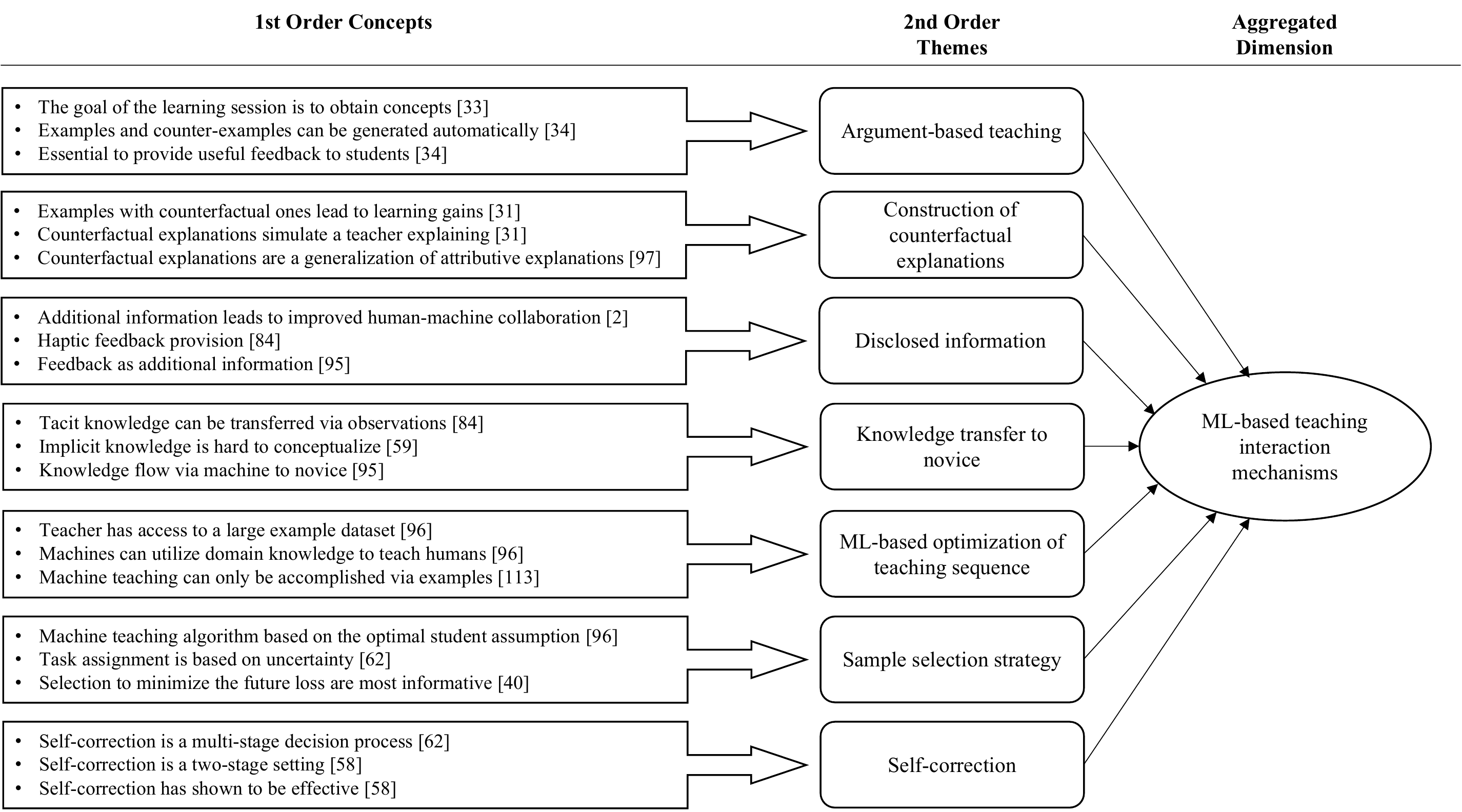}}
	\caption{Data structure of ML-based teaching interaction mechanisms.}
	\label{datastructure_teim}       
\end{figure}

ML-based teaching systems incorporate different interaction mechanisms between ML-based teachers and novices  (\Cref{datastructure_teim}). These are established to enable a \textbf{knowledge transfer} to novices for specific tasks. Compared to traditional intelligent tutoring systems, where domain knowledge must be well-structured, ML-based teaching systems can use their ability to learn from data samples and derive knowledge. \citet{stein2013machines} argue that SMEs cannot easily articulate tacit knowledge and that acquiring expert knowledge, in general, is challenging. This is why ML models are well suited to derive that knowledge from data samples. 

In \textbf{argument-based} machine learning (ABML) approaches, knowledge is transferred via arguments SMEs provide in learning examples \citep{guid2019automated}. The authors in \citet{guid2019automated} emphasize the ability of such mechanisms to provide automatic examples that novices are assigned to explain. One important aspect is providing feedback to novices so they can learn new concepts. \citet{zapuvsek2014designing} highlight the capability of such an interaction to draw expert knowledge from domains and enable a \enquote{powerful knowledge elicitation tool} \citep[p. 575]{zapuvsek2014designing}. 

MT as an interaction mechanism of \textbf{optimizing teaching sequences} focuses on providing optimal teaching sets to novices \citep{zhu2018overview}. This is inspired by conventional teaching \citep{mac2018teaching}. The interaction between ML-based teacher and novice in MT-based teaching systems is, in most cases, restricted to the provision of examples \citep{zhu2013machine}. Here, the \textbf{sample selection strategy} can build on assumptions of cognitive models of novices \citep{zhu2015machine, zhang2020interactive}. \citet{zhou2018unlearn}, for instance, use the decay memory model of novices. In \citet{patil2014optimal}, the authors base the sample selection of their teaching system on a limited capacity model and justify this choice with the cognition capabilities of humans. \citet{johns2015becoming} select the teaching examples dependent on the novice's knowledge. 

In general, the ML model in ML-based teaching systems represents an intermediary to facilitate the transfer of task-specific knowledge from SMEs to novices \citep{wang2019extracting}. In addition to presenting examples to the novice, in some teaching systems, the teaching sequence is optimized through revealing \textbf{additional information}. Such explanations, which can have the form of arguments and counterexamples \citep{guid2019automated}, can also be displayed to the novice next to the ML models' predictions. These refer to the answer estimated by the novice \citep{matsubara2018learning}. This form of interaction can lead to \textbf{self-correction} and has shown to be effective in teaching novices \citep{nakayama2021crowd}. As stated in \Cref{basic_elements_section}, the level of uncertainty can be disclosed to novices as another form of explanation to have them reconsider their answer \citep{abdel2020and}. In \citet{mac2018teaching}, the authors state the mere provision of examples with their ground truth as limited feedback. Instead, the authors argue that interpretable feedback will lead to a better learning process. \citet{su2017interpretable} express the same opinion by outlining that instructions shall be clear and interpretable. In \citet{stein2013machines}, the authors describe a teaching system in which the interaction between novice and ML-based teachers is provided through haptic feedback. This real-time feedback enables the novice to learn how to steer a crane.

As form of presenting examples with explanations to novices, counterfactual explanations provide reasoning for \enquote{why a mistake was made} \citep[p. 8983]{wang2020scout}. \citet{goyal2019counterfactual} base the \textbf{construction of counterfactual explanations} on discriminative regions. They argue that such explanations reveal the differences between two examples. These explanations result in increased knowledge gain. Moreover, \citet{wang2020scout} claim that counterfactual explanations can explain why a mistake was made and hence, provide means to learning a task.

\subsection{ML Training Interaction Mechanisms}
\begin{figure}[htbp!]
    \centering{\includegraphics[clip, scale=0.53]{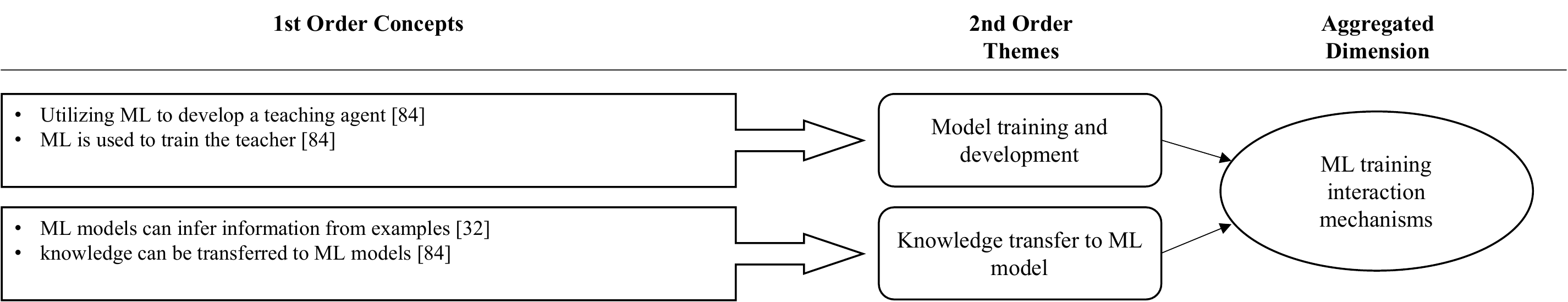}}
	\caption{Data structure of ML training interaction mechanisms.}
	\label{datastructure_trim}       
\end{figure}

As ML models facilitate the teaching process between an artificial teacher and a human novice, the \textbf{development} and training phase require some form of interaction with an SME to build the models  (\Cref{datastructure_trim}). This development depends on the type of ML model used in the teaching system. The choice mainly depends on the functions that the ML model is taking over in the teaching procedure. For providing explanations, for instance, \citet{goyal2019counterfactual} utilize convolutional neural networks to highlight relevant features on images. \citet{singla2014near} make use of a Markov chain model to select examples for the novice. 

The kind of support ML models provide varies across domains and requires training to sustain a \textbf{knowledge transfer} from subject matter experts towards ML models. Since SMEs cannot articulate their tacit knowledge, they draw on examples to pass the expert knowledge towards ML models \citep{zhu2018overview}. Thus, ML models can infer information or structures from such examples and form their understanding of the domain \citep{stein2013machines}. This advantage of passing on the knowledge via examples is superior to traditional intelligent tutoring systems, which relies on well-structured domain knowledge to be set up \citep{gross2015learning}.

\subsection{Teaching Reflection Mechanisms}
\begin{figure}[htbp!]
    \centering{\includegraphics[clip, scale=0.53]{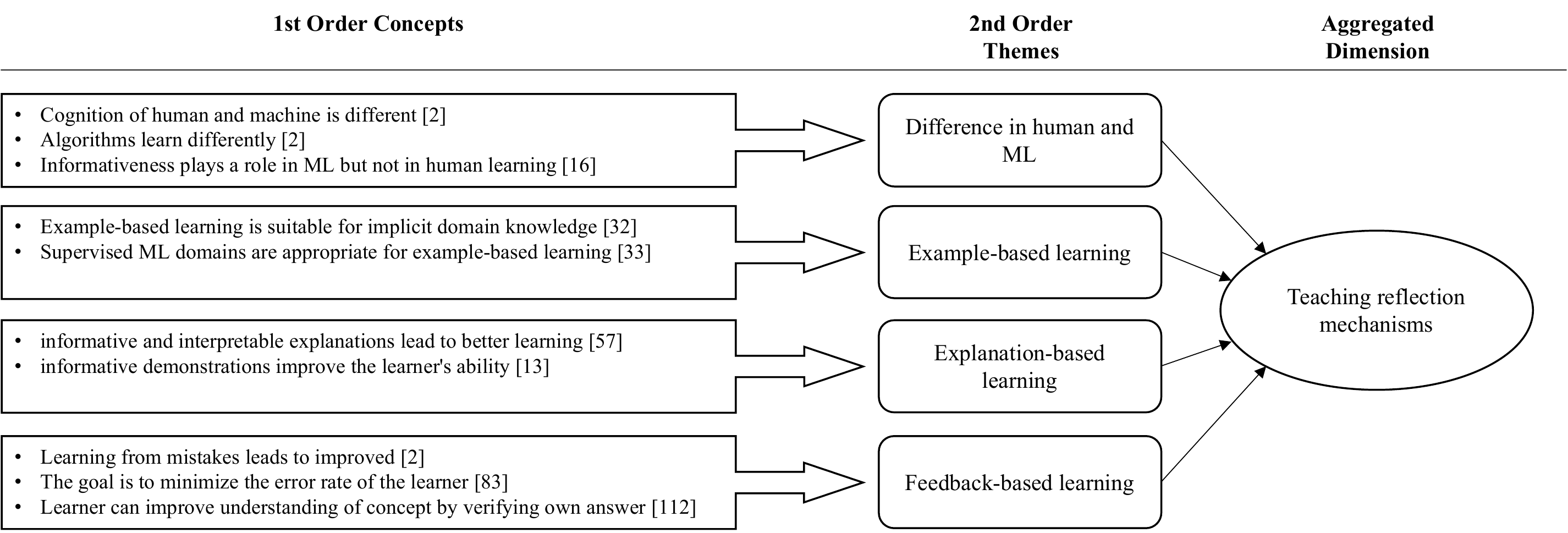}}
	\caption{Data structure on teaching reflection mechanisms.}
	\label{datastructure_rm}       
\end{figure}

Teaching reflection mechanisms define the processes which lead to the formation of novice's knowledge (\Cref{datastructure_rm}). The teaching systems identified in our literature review are set up as \textbf{example-based learning} systems. In such a workflow, novices are provided data samples of the domain and asked to perform a task on those samples. \citet{wang2019extracting} show the effectiveness of this approach by utilizing a teaching system to transfer skills through demonstrations. Especially supervised learning domains are appropriate for example-based learning procedures since labeled samples are available \citep{guid2019automated}.

Next to the mere provision of samples, disclosed information can enhance the learning process of the novice, as stated in the previous section. In this \textbf{explanations-based learning}, novices comprehend a concept better \citep{goyal2019counterfactual}. However, \citet{chen2018near} state that the information of an example is only useful to a learner if they can access it. Thus, informative and interpretable explanations lead to a better understanding by novices \citep{mac2018teaching}.

While explanations can be provided directly to the novice, \textbf{feedback-based learning} is established by reacting to a novice's answer in providing additional information. The goal of such feedback is to reduce the errors of a learner \citep{singla2014near}. Through receiving additional information after making a mistake, novices can improve their knowledge by revising their initial conception of the domain\citep{abdel2020and}.

By analyzing different reflection mechanisms, we revealed another concept of ML-based teaching systems. While ML models are at the core of such teaching systems, the learning process of these models \textbf{differs} from the one of humans. \citet{patil2014optimal} state the difference by comparing the training procedures on non-representative examples. Human learners are more sensitive toward specific features of samples and cannot perform a task optimally just by observing examples. This has implications for teaching systems since the samples used to train an ML model are not the best examples to provide to a human learner \citep{basu2013teaching}.

\section{A Conceptualization of ML-based Teaching Systems}
\label{framework_section}
\begin{figure}[htbp!]
    \centering{\includegraphics[clip, scale=0.46]{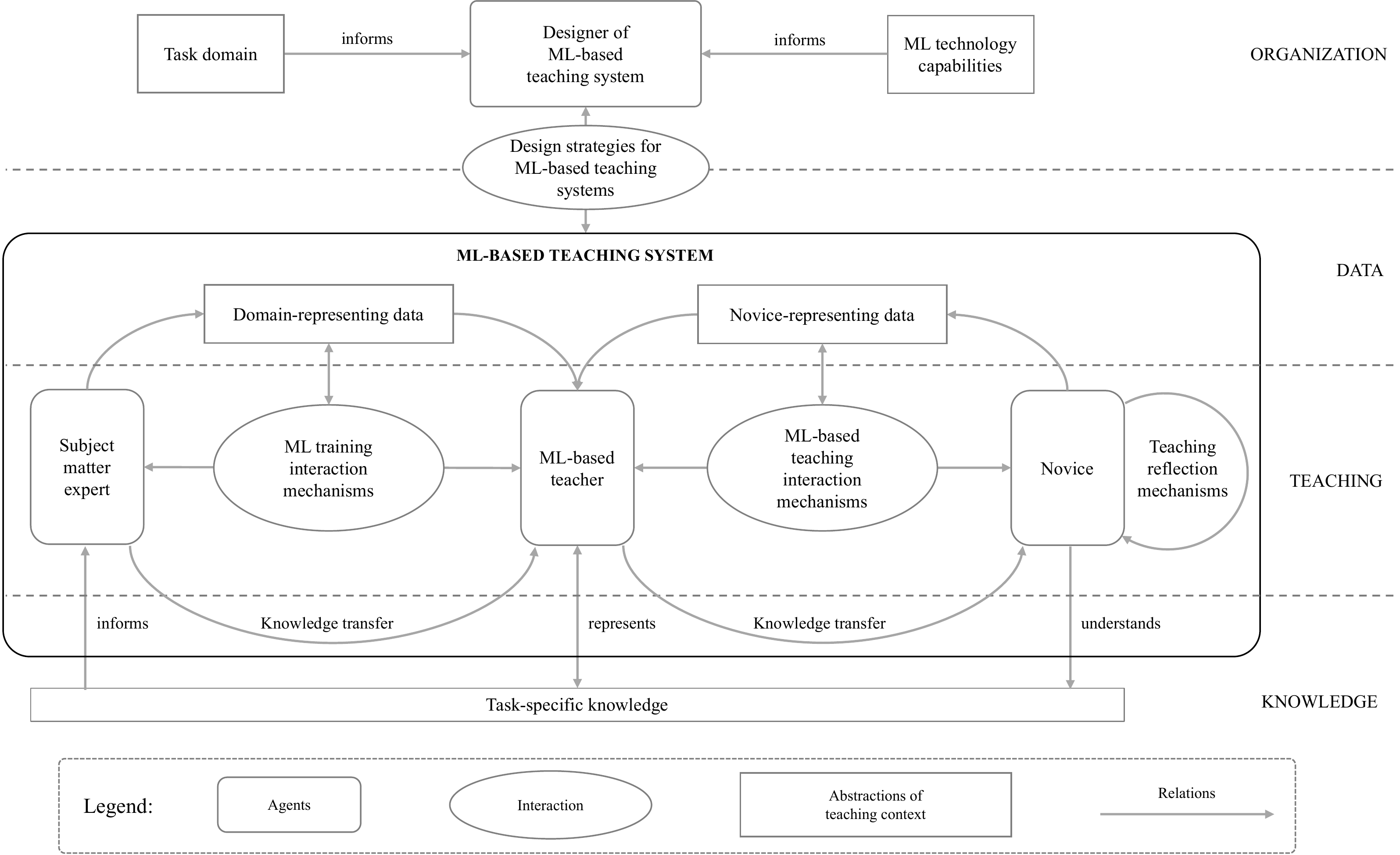}}
	\caption{Conceptual framework of ML-based teaching systems.}
	\label{framework}       
\end{figure}

Regarding our research question on how ML can facilitate the transfer of knowledge in organizational settings, we structure our SLR results and conceptualize our findings in a framework. As mentioned in \Cref{methodology_section}, we use a methodology to articulate the grounded theory that we synthesized in the previous section \citep{gioia2013seeking}. Thus, we outline our findings in constituting the relations of identified second order themes and aggregated dimensions. The respective framework of ML-based teaching systems is illustrated in \Cref{framework}.

Four dimensions are essential in ML-based teaching systems: Organization, data, teaching, and knowledge. The latter represents the core of such teaching systems, as the overall goal is to transfer knowledge from SMEs to novices. One distinction to conventional teaching, in which the expert trains the novices directly (\Cref{teaching}), is the indirect knowledge transfer in ML-based teaching systems. Ml-based teachers are an intermediary between SME and novice \citep{wang2019extracting}. There are two mechanisms that this intermediary function is grounded on. The first is an ML interaction mechanism describing an ML-based teacher's building phase. Here, the focus is on training an ML model. The second is the ML-based teaching interaction mechanism that defines how the ML-based teacher interacts with the novice. Such mechanisms can be comprised of presenting examples or explanations to the novice. In addition, the novice is subject to reflection mechanisms leading to comprehending concepts and enabling them to understand the task.

SMEs reveal their knowledge in the form of data that composes the link of knowledge transfer between SME and ML-based teachers. Besides such domain-representing data, novices generate data based on their answers in the teaching process. For instance, novices might provide the prediction on the house price evaluation task. The ML-based teacher can process these answers to provide informative feedback to the novice. 

A human designer can specify the construction of ML-based teaching systems by optimally adopting the system's design according to the domain's requirements. Here, knowledge of different ML capabilities is crucial to establish a compatible teaching system.

\section{Discussion}
\label{discussion_section}
The ongoing need to retain expert knowledge and transfer it has increased the demand for alternative solutions to teach novices. Yet, a conceptual understanding of the properties and capabilities of ML-based teaching systems remains limited and distributed across disciplines, and human-computer interaction literature misses synthesizing existing insights of such systems to be able to extend them. Therefore, we set out to answer how ML can facilitate the knowledge transfer in organizational teaching. Based on a sample of 29 scientific articles, this work presents the findings of an inductive review on ML-based teaching systems, which uses a grounded theory-inspired method for rigorously reviewing literature \citep{wolfswinkel2013using}. In doing so, we distill relevant concepts in the literature sample, which we iteratively aggregate into 28 themes in research on ML-based teaching systems. Further, we identify and develop relations between these main themes that lead us to 5 aggregated dimensions, which we discuss in \Cref{result_section} in detail. This three-step coding process allowed us also to re-conceptualize the existing understanding of what an ML-based teaching system is in form of a conceptual framework, which is presented in \Cref{framework_section}.

\subsection{Theoretical Implications and Emerging Research Avenues}
The research we undertook that led to this conceptualization provides at least four sets of implications for research, namely the role of the organization, data, teaching, and knowledge; dimensions that already helped us to structure our conceptual framework. In this section, we summarize these sets of implications and identify nine research avenues that emerged from our study and which we discuss in turn.

First, our review on ML-based teaching systems sheds new light on the \textbf{role of the organization} in the design of ML-based teaching systems. The organizational context, and in particular the task domain as well as the ML technological capabilities recognized by the organization, form the basis for deciding between different design strategies and thus for the implementation of ML-based teaching systems. For a successful system design, it is therefore crucial to determine the requirements of this socio-organizational context beforehand. Previous research has already investigated the requirements for ML-based teaching systems from a novice's perspective \citep{qin2020understanding, kim1, kim2} or an ML model point of view \citep{johns2015becoming, wang2020scout}. However, the literature lacks a holistic view of the broad and nuanced possible differentiations of the organizational context that lead to different requirements and perceptions that guide the design of ML-based instructional systems. Thus, we identify the following two emerging research avenues:
\begin{avenue}
    \item How do differences in the task domain of an organization translate into different requirements for the design of ML-based teaching systems?
\end{avenue}

\begin{avenue}[resume]
    \item How can organizations turn the general technological capabilities presented by machine learning into an effective design of ML-based teaching systems in their organizational context?
\end{avenue}

To apply the framework in practice and provide guidance to knowledge managers, it is necessary for designers of ML-based teaching systems to determine work practices and requirements for the successful implementation of such systems \citep{schmidt2011cooperative}. As designers operate at the intersection of task domain and ML capabilities (\Cref{framework}), related work on ML pipelines \citep{shang2019democratizing} and ML operations \citep{kreuzberger2023machine} for data science projects can be a starting point for determining such practices. Furthermore, \citet{stumpf2009interacting} explore user interaction and trust in machine learning systems. Similarly, the  perception and trust of novices in ML-based teaching systems in organizational settings need. Recent research in CSCW already investigates work practices, for instance, of data scientists \citep{muller2019human} and machine learning developers \citep{wolf2019conceptualizing} in organizational contexts, or more specifically, machine teachers (with a focus on the teacher's interaction with data \citep{simard2017machine} and the IT specialist) in the customer service domain \citep{candello2022unveiling}. Therefore, articulation work and work practices must be determined for the development of ML-based teaching systems in an organizational context:

\begin{avenue}[resume]
    \item What are work practices of designers and novices of ML-based teaching systems in an organizational context?
\end{avenue}

Additionally, several stakeholders are involved in the deployment of ML-based teaching systems. For instance, the designers of such systems, as shown in \Cref{framework}, need to examine the task domain to determine relevant content and requirements for the teaching system to be successful. Furthermore, as research on human-data interaction shows, data stakeholders play a crucial role in data-based services \citep{seidelin2018data}. With multiple stakeholders involved in the development and use of ML-based teaching systems, we derive the following research avenue:

\begin{avenue}[resume]
    \item How can designers of ML-based teaching systems translate stakeholders' needs and requirements into successful implementation?
\end{avenue}


Second, by specifying domain-representing data and novice-representing data as two distinct types of relevant data in ML-based teaching systems, our results provide a more nuanced view of the \textbf{role of data} in such systems. While the former is mainly used to formalize the interaction between the SME and the ML model, the latter is used as input to control and optimize the teaching interaction with the novice over time and ultimately to a formalized understanding of the teaching context. Future research can adopt this differentiation to be more precise about the data generated in their specific teaching context, and how it was used to train an ML-based teacher. 
Given their intangible nature, such datasets might also be reused in multiple other ML-based teaching systems within and beyond the organization. Future research could therefore pursue the following emerging research avenue:

\begin{avenue}[resume]
    \item How can datasets resulting from interactions in ML-based teaching systems be systematically reused in other teaching contexts within and beyond the organization?
\end{avenue}

In the work of \citet{johns2015becoming} and \citet{mac2018teaching} the authors highlight that instructions and explanations need to be considered thoroughly and shall be interpretable and informative to ensure a sufficient teaching process. However, for an ML-based teaching system developed on vast amounts of sensitive data and to be deployed in organizational contexts that aim to qualify novices as crucial assets, considerations of security \citep{sahay2021advances} need to be taken into account. \citet{zhu2018overview} describe the difference of honest and dishonest Ml-based teachers. Considering the sensitive context in which teaching systems are used and the evolving research on ML-based teaching systems, requirements for safe teaching need to be determined:
\begin{avenue}[resume]
    \item What are requirements for secure ML-based teaching systems in organizational settings?
\end{avenue}

Third, at the core of our conceptualization, three types of mechanisms emerge that systematize the elementary \textbf{role of teaching} manifested in ML-based teaching systems: 1) The interaction between SME and ML-based teacher to train the teacher, 2) the interaction between the ML-based teacher and the novice to teach the novice, and 3) the novice's self-reflection of the teaching experience. While our review collects a variety of approaches to bring each type of mechanism to life, we would like to highlight certain themes that seem fruitful for future research.

An aspect suggested by our results is that different ML approaches differ, particularly in how they can transfer knowledge \citep{kane2009shoemaker, atapattu2015educational}. ML systems can be enhanced as an intermediary to capture not only the explicit part of knowledge but also its tacit form by learning from data \citep{fenstermacher2005tyranny, stein2013machines}. However, we do not yet know which ML approaches should be preferred for learning and teaching which tasks  \citep{abdel2020and}, and how to design machine-in-the-loop\footnote{Machine-in-the-loop systems require a machine to interact with a human to influence the human behaviour in specific actions} processes in each case for optimal training of an ML-based teacher in interaction with an SME. Thus, we derive the need to review the interaction of SME and ML-based teachers in this context:
\begin{avenue}[resume]
    \item Which ML approaches should be preferred for learning and teaching which tasks?
    \item How should machine-in-the-loop processes be designed for an optimal training of an ML-based teacher in interaction with an SME?
\end{avenue}

With the rising use and capabilities of ML-based teaching systems, the literature points out that perception of novices in the interaction with ML models \citep{kim1}, trust \citep{qin2020understanding} but also communication styles between ML models and novices \citep{kim2} have a crucial impact on teaching success. This emphasizes the need to investigate further how this training can advance future collaborations of ML models and humans by teaching the novice early in the teaching process how to cooperate with ML systems:
\begin{avenue}[resume]
    \item How do ML-based teaching interaction episodes impact potential future interactions between novices and ML-based teachers?
\end{avenue}

In terms of the self-reflection of the novice, we recommend further investigation into how techniques related to explainable artificial intelligence (XAI) may influence the continuation of a novice's learning experience while taking a psychological point of view. For example, \citet{goyal2019counterfactual} and \citet{wang2020scout} show how XAI can affect the learning progress of novices. In this context, it can be crucial to consider human traits on a personal but also cognitive level to reveal how different explanations and instructions of XAI affect the stimuli of novices and their ability to comprehend knowledge:
\begin{avenue}[resume]
    \item How do cognitive and individual preferences of novices affect the stimuli of explanations by XAI and impact the ability to learn through ML-based teaching systems?
\end{avenue}

Finally, we position task-specific knowledge as the linking element for the interacting agents in ML-based teaching systems, and therefore emphasize the integrative \textbf{role of knowledge}. While their task-specific knowledge informs SMEs to interact within the teaching system and novices improve their understanding over time, we position the ML-based teacher as a new representative of task-specific knowledge. While the representation of task-specific knowledge on visual classification tasks has been investigated many times \citep{castro2008human, johns2015becoming, su2017interpretable, mac2018teaching}, representing other kinds of task-specific knowledge such as on regression-alike tasks are not yet sufficiently examined. Thus, we point out the following emergent research avenue:
\begin{avenue}[resume]
    \item What kinds of task-specific knowledge is an ML-based teacher able to represent and why?
\end{avenue}

\subsection{Managerial Implications}
This work presents important managerial implications by reflecting on existing ML-based teaching systems and drawing insights for the design and application of such systems to transfer expert knowledge in an organization. Our conceptualization helps to adopt a more systematic comprehension to design and implement such systems at the intersection of ongoing business operations and knowledge retention. In addition, we address multiple stakeholders with our framework: Knowledge managers to guide knowledge retention activities, designers to successfully develop and implement ML-based teaching systems at the intersection of task domain and ML capabilities and novices as end-users of ML-based teaching systems. We encourage practitioners designing ML-based teaching systems to apply our conceptual framework to structure their design approach. Our review on existing articles might further point out applicable knowledge on the respective design areas of such systems. Moreover, this can aid organizations in laying out the strategic direction of knowledge retention. With the proposed research avenues we link ML-based teaching systems to related fields that might encourage the transfer of knowledge. These avenues point out important linkages between the development of ML-based teaching systems and the adoption in organizations which are of interest not only to research but also to practice.

\subsection{Limitations}
Our study certainly comes with some limitations. By searching selected databases we limit the scope of the SLR to its full extent. This might delimitate the generalizability of our results. Additionally, the articles analyzed in this work are based on and therefore restricted by our search term. Finally, there is a risk of human error in a literature search, even though we followed established methods. We carried out workshops between three researchers with the aim of a shared understanding and coherent results to minimize that risk \citep{hund2021digital}. The research avenues we derived from our findings are based on only theoretical foundations. This highlights the need to further examine this topic with empirical research approaches. Therefore, these limitations present a fruitful starting point for future research on ML-based teaching systems. 

\section{Conclusion}
\label{conclusion_section}
This article sets out an inductive review on ML-based teaching systems by analyzing and conceptualizing how these can be leveraged in organizational settings. So far, human-computer interaction and CSCW literature lacks to thoroughly scrutinize a common understanding of the capabilities of ML-based teaching systems to facilitate knowledge transfer. Hence, through a qualitative analysis based on grounded theory we inductively reviewed literature to reveal how existing research scrutinizes ML-based teaching systems and made two contributions: First, by identifying relevant articles in the domain we obtained essential concepts for ML-based teaching systems and conceptualized a framework to synthesize an understanding for their links and interrelations. In doing so, we reveal how ML can facilitate the knowledge transfer in organizational teaching. Second, based on our findings, we propose a future research agenda with several research avenues. 

Overall, with this work we intend to reveal functional and structural capabilities of ML models in teaching systems and to better understand how they can be leveraged to support knowledge transfer in organizational settings. Extensive and rigorous research is needed to fully understand and exploit ML-based teaching systems. We invite researchers to take part in this debate and hope to inspire scientists to actively participate in this endeavor --- so that we can finally start imagining: Through ML-based teaching, your chess game has undoubtedly improved and you have left your best friend speechless. Next time, they will ask you to teach them.

\bibliographystyle{ACM-Reference-Format}
\bibliography{sample-base}

\newpage
\appendix
\section{Appendix}
\label{appendix}

\begin{table*}[htbp!]
\caption{Concept matrix of aggregated dimensions.}
\centering
\begin{tabular}{m{3.5cm} P{1.9cm} P{1.9cm} P{1.9cm} P{1.9cm} P{1.9cm}}
\toprule
\multirow{2}{*}{Article} & \multicolumn{5}{c}{Aggregated Dimensions}                           \\ \cmidrule{2-6}
                         & Basic Elements of ML-based Teaching Systems & Design Strategies for ML-based Teaching Systems & ML-based Training Interaction Mechanisms & ML-based Teaching Interaction Mechanism & Teaching Reflection Mechanisms \\
\midrule
\citet{abad2017autonomous} & \ding{53} & \ding{53}  & \ding{53} &   &   \\
\citet{abdel2020and} & \ding{53} & \ding{53}  & \ding{53}  &  & \ding{53}  \\
\citet{basu2013teaching} & \ding{53} & \ding{53} & \ding{53} & \ding{53}  & \ding{53} \\
\citet{cakmak2012algorithmic} & \ding{53} & \ding{53} & \ding{53}  & \ding{53}  & \ding{53} \\
\citet{castro2008human} & \ding{53} & \ding{53}  & \ding{53}  &  &   \\
\citet{chen2018near} & \ding{53} & \ding{53} & \ding{53} & \ding{53}  & \ding{53}\\
\citet{goyal2019counterfactual} & \ding{53} &  \ding{53} & \ding{53} &   &   \\
\citet{gross2015learning} & \ding{53} & \ding{53}  & \ding{53} & \ding{53}  & \ding{53}  \\
\citet{guid2019learning} & \ding{53} & \ding{53}  & \ding{53} &   &  \\
\citet{guid2019automated} & \ding{53} & \ding{53}  & \ding{53} &   & \ding{53}  \\
\citet{johns2015becoming} & \ding{53} & \ding{53}  & \ding{53} & \ding{53}  &  \\
\citet{mac2018teaching} & \ding{53} &  & \ding{53} & \ding{53}  & \ding{53} \\
\citet{matsubara2018learning} & \ding{53} &  & \ding{53} &   &  \\
\citet{movzina2012goal} & \ding{53} & \ding{53} & \ding{53} &   &  \\
\citet{nakayama2021crowd} & \ding{53} & \ding{53} & \ding{53} &   &  \\
\citet{patil2014optimal} & \ding{53} & \ding{53} & \ding{53} & \ding{53}  & \ding{53} \\
\citet{singla2014near} & \ding{53} & & \ding{53} & \ding{53}  & \ding{53} \\
\citet{stein2013machines} & \ding{53} & \ding{53} & \ding{53} & \ding{53}  &  \\
\citet{su2017interpretable} & \ding{53} & \ding{53} & \ding{53} & \ding{53}  & \ding{53} \\
\citet{wang2019extracting} & \ding{53} & \ding{53} &  & \ding{53}  & \ding{53} \\
\citet{wang2021gradient} & \ding{53} & \ding{53} & \ding{53} & \ding{53}  & \\
\citet{wang2020scout} & \ding{53} & \ding{53} & \ding{53} & \ding{53}  & \\
\citet{wang2021machine} & \ding{53} & \ding{53} & \ding{53} & \ding{53}  & \ding{53} \\
\citet{zapuvsek2014designing} &  & \ding{53} & \ding{53} &   & \\
\citet{zhang2020interactive} & \ding{53} & \ding{53}  & \ding{53} &   &  \\
\citet{zhou2018unlearn} & \ding{53} & \ding{53} &  & \ding{53}  & \ding{53}\\
\citet{zhu2013machine} & \ding{53} &  & \ding{53} & \ding{53}  & \\
\citet{zhu2015machine} & \ding{53} &  & \ding{53} &   & \\
\citet{zhu2018overview} & \ding{53} & \ding{53} & \ding{53} & \ding{53}  & \\
\midrule
\\
\end{tabular}

\label{literaturematrix}
\end{table*}

\end{document}